\documentstyle[psfig]{mn}

\newif\ifAMStwofonts

\newcommand{\lapprox}{\stackrel{<}{\scriptstyle \sim}}
\newcommand{\gapprox}{\stackrel{>}{\scriptstyle \sim}}

\ifoldfss
  \ifCUPmtlplainloaded \else
    \NewTextAlphabet{textbfit} {cmbxti10} {}
    \NewTextAlphabet{textbfss} {cmssbx10} {}
    \NewMathAlphabet{mathbfit} {cmbxti10} {} 
    \NewMathAlphabet{mathbfss} {cmssbx10} {} 
  \fi
  \ifAMStwofonts
    \ifCUPmtlplainloaded \else
      \NewSymbolFont{upmath} {eurm10}
      \NewSymbolFont{AMSa} {msam10}
      \NewMathSymbol{\upi}     {0}{upmath}{19}
      \NewMathSymbol{\umu}     {0}{upmath}{16}
      \NewMathSymbol{\upartial}{0}{upmath}{40}
      \NewMathSymbol{\leqslant}{3}{AMSa}{36}
      \NewMathSymbol{\geqslant}{3}{AMSa}{3E}

       \let\le=\leqslant
       \let\ge=\geqslant
    \fi
  \fi
\fi 

\ifnfssone
  \newmathalphabet{\mathit}
  \addtoversion{normal}{\mathit}{cmr}{m}{it}
  \addtoversion{bold}{\mathit}{cmr}{bx}{it}
  \newmathalphabet{\mathbfit} 
  \addtoversion{normal}{\mathbfit}{cmr}{bx}{it}
  \addtoversion{bold}{\mathbfit}{cmr}{bx}{it}
  \newmathalphabet{\mathbfss} 
  \addtoversion{normal}{\mathbfss}{cmss}{bx}{n}
  \addtoversion{bold}{\mathbfss}{cmss}{bx}{n}
  \ifAMStwofonts
    \ifCUPmtlplainloaded \else
      \UseAMStwoboldmath
      \makeatletter
      \new@mathgroup\upmath@group
      \define@mathgroup\mv@normal\upmath@group{eur}{m}{n}
      \define@mathgroup\mv@bold\upmath@group{eur}{b}{n}
      \edef\UPM{\hexnumber\upmath@group}
      \new@mathgroup\amsa@group
      \define@mathgroup\mv@normal\amsa@group{msa}{m}{n}
      \define@mathgroup\mv@bold\amsa@group{msa}{m}{n}
      \edef\AMSa{\hexnumber\amsa@group}
      \makeatother
      \mathchardef\upi="0\UPM19
      \mathchardef\umu="0\UPM16
      \mathchardef\upartial="0\UPM40
      \mathchardef\leqslant="3\AMSa36
      \mathchardef\geqslant="3\AMSa3E

       \let\le=\leqslant
       \let\ge=\geqslant
    \fi
  \fi
\fi 

\ifnfsstwo
  \DeclareMathAlphabet{\mathbfit}{OT1}{cmr}{bx}{it}
  \SetMathAlphabet\mathbfit{bold}{OT1}{cmr}{bx}{it}
  \DeclareMathAlphabet{\mathbfss}{OT1}{cmss}{bx}{n}
  \SetMathAlphabet\mathbfss{bold}{OT1}{cmss}{bx}{n}
  \ifAMStwofonts
    \ifCUPmtlplainloaded \else
      \DeclareSymbolFont{UPM}{U}{eur}{m}{n}
      \SetSymbolFont{UPM}{bold}{U}{eur}{b}{n}
      \DeclareSymbolFont{AMSa}{U}{msa}{m}{n}
      \DeclareMathSymbol{\upi}{0}{UPM}{"19}
      \DeclareMathSymbol{\umu}{0}{UPM}{"16}
      \DeclareMathSymbol{\upartial}{0}{UPM}{"40}
      \DeclareMathSymbol{\leqslant}{3}{AMSa}{"36}
      \DeclareMathSymbol{\geqslant}{3}{AMSa}{"3E}

       \let\le=\leqslant
       \let\ge=\geqslant
    \fi
  \fi
\fi 

\ifCUPmtlplainloaded \else
  \ifAMStwofonts \else 
    \def\upi{\pi}
    \def\umu{\mu}
    \def\upartial{\partial}
  \fi
\fi

\title[Field Statistics for Vela's Pulsar: 2]
{Intrinsic Variability and Field Statistics for the Vela Pulsar: 2. Systematics 
and Single-Component Fits}
\author[Cairns, Johnston and Das] 
       {Iver~H. Cairns, S. Johnston, and P. Das\\
        School of Physics, University of Sydney, NSW 2006, Australia.}
\date{Accepted ??.
      Received 2002 ??;
      in original form 2002 ??}

\pagerange{\pageref{firstpage}--\pageref{lastpage}}
\pubyear{2002}

\begin{document}

\maketitle

\label{firstpage}

\begin{abstract}
Individual pulses from pulsars have 
intensity-phase profiles that differ widely from pulse to pulse, 
from the average profile, and from phase to phase within a pulse. 
Widely accepted explanations do not exist for this variability or 
for the mechanism producing the radiation. The variability 
corresponds to the field statistics, particularly the 
distribution of wave field amplitudes, which are predicted by theories for 
wave growth in inhomogeneous media. This paper shows that the field 
statistics of the Vela pulsar (PSR B0833-45) are well-defined and 
vary as a function of pulse phase, evolving from Gaussian intensity statistics 
off-pulse to approximately power-law and then lognormal distributions near the pulse 
peak to approximately power-law and eventually Gaussian statistics 
off-pulse again. Detailed single-component fits confirm that the 
variability corresponds to lognormal statistics near the peak of the 
pulse profile and Gaussian intensity statistics off-pulse. The  
lognormal field statistics observed are consistent with the 
prediction of stochastic growth theory (SGT) for a purely linear 
system close to marginal stability. The simplest interpretations are that the 
pulsar's variability is a direct manifestation of an SGT state and the 
emission mechanism is linear (either direct or indirect), 
with no evidence for nonlinear mechanisms like modulational instability and 
wave collapse which produce power-law field statistics. Stringent 
constraints are  placed on nonlinear mechanisms: they must produce lognormal 
statistics when suitably ensemble-averaged. 
Field statistics 
are thus a powerful, potentially widely applicable tool for understanding 
variability and constraining mechanisms and source characteristics 
of coherent astrophysical and space emissions. 
 
\end{abstract}


\section{Introduction}

  Pulsars are believed to be highly magnetized neutron stars whose 
rotation causes highly nonthermal beams of radiation to be swept 
across the Earth. Most 
likely the radiation is produced over the magnetic polar caps of the star, 
which are offset 
from the rotational poles. The radiation's 
high brightness temperatures require coherent emission processes such as 
plasma microinstabilities or nonlinear processes; however, despite many years 
of research, no agreement exists on which mechanisms dominate  
\cite{asseo1996,hankins1996,melrose1996,melrosegedalin1999}. Proposed 
linear 
instabilities include:  
(i) Linear acceleration and maser curvature emission 
\cite{luomelrose1995,melrose1996},  in which 
electrons radiate coherently while accelerating in an oscillating 
large-scale field or on curved magnetic field lines, respectively. 
(ii) Relativistic plasma emission \cite{melrose1996,asseo1996}, in which 
a streaming instability 
either directly generates escaping radiation near harmonics of the 
electron plasma frequency $f_{pe}$ or else drives localized, non-escaping  
waves near $f_{pe}$ that are converted into escaping harmonic radiation by 
linear mode conversion or nonlinear processes. (iii) A streaming instability 
into a new, directly escaping mode (Gedalin, Gruman \& Melrose 2002). Possible nonlinear 
mechanisms involve solitons, modulational instabilities, and strong turbulence 
wave collapse of intense localized wavepackets of waves driven near 
$f_{pe}$ (Pelletier, Sol, \& Asseo 1988, Asseo, Pelletier, \& Sol 1990, 
Asseo 1996, Weatherall 1997, 1998).  
Another 
possibility is an antenna mechanism, in which largescale low 
frequency wavepackets act as antennas for conversion of other higher 
frequency waves into radiation, due to acceleration of electrons in the 
combined wave fields 
(Pottelette, Treumann \& Dubouloz 1999, Cairns \& Robinson 2001).
Standard analyses 
of linear and nonlinear growth rates suggest that numerous mechanisms are 
viable, in part due to uncertainties in the source plasma characteristics 
and location of emitting regions (e.g., above the polar cap or near the 
light cylinder). Accordingly new approaches are necessary. 

\begin{figure}
\psfig{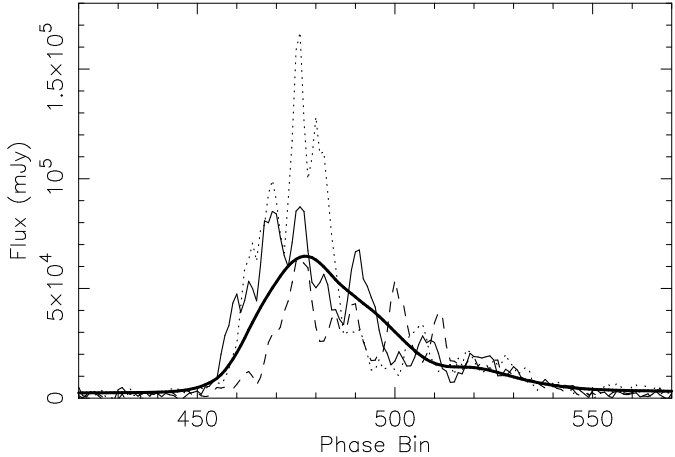}
\caption{Average flux profile as a function of phase bin for the 
Vela dataset (thick line), together with three superposed individual 
pulses (other lines). See Section 3 for more details.}
\label{fig1}
\end{figure}

    Since the  discovery of pulsars it has been known that only suitably long time 
averaging leads to a stable intensity profile. 
While this average profile is unique to each pulsar, individual pulses vary widely in 
intensity, often by a factor of 5 or more, from one phase to another in 
a given pulse and from one pulse to the next at a given phase. Illustrated 
in Figure \ref{fig1} for Vela, this variability is intrinsic, clearly 
related to the emission mechanism and/or propagation effects, 
and has no accepted interpretation. This paper considers 
the variability in terms of its statistics, following Cairns, Johnston \& 
Das (2001),hereafter called Paper I. 
In general, the variability includes phenomena known as drifting  
subpulses, microstructures, giant pulses, and giant micropulses.
Subpulses are features that drift in time across the pulse window 
\cite{drakecraft1968,manchestertaylor1977}, while microstructures 
are concentrated features superposed on a subpulse that are sometimes 
quasiperiodic 
(Craft, Comella \& Drake 1968, Kramer et al. 2002). 
Giant pulses
\cite{cognardetal1996,hankins1996} and giant micropulses \cite{johnstonetal2001} 
are very rare pulses with pulse-integrated fluxes $\gapprox 10$ times the average. 

   Analyses of field statistics, such as the distributions of electric field 
strengths or intensities, are not yet 
standard in analyses of astrophysical radiation, in contrast to Fourier 
and correlation analyses related to propagation effects \cite{rickett1977}. 
Probable reasons include (i) the strong theoretical motivations and benefits of 
considering field statistics were not clear before the advent of 
stochastic growth theory (SGT) 
(Robinson 1992, 1995, Robinson, Cairns \& Gurnett 1993, 
Cairns, Robinson \& Anderson 2000, Robinson \& Cairns 2001, Cairns \& Menietti 
2001) 
and other theories such as self-organized criticality [SOC] 
(Bak, Tang \& Weisenfeld 1987, Bak 1996), 
(ii) these motivations and benefits are not widely known 
in the astrophysical community, and (iii) high time resolution, 
coherently dedispersed data for high intensity sources, whose statistics are not 
strongly contaminated by instrumental/background noise, have only recently 
become available \cite{johnstonetal2001,vanstratenetal2001,krameretal2002}. 
Recent analyses of seven solar system wave phenomena show that all have 
well-defined field statistics that agree very well with the functional 
forms predicted by SGT, resolving longstanding theoretical problems pertaining to
the burstyness, widely varying fields, and persistence of the waves (and 
driving particles) for unexpectedly large distances in the source. 
Giant pulses of some pulsars have power-law flux distributions 
\cite{cognardetal1996}, as do giant micropulses 
\cite{johnstonetal2001}, potentially being interpretable in 
terms of SOC \cite{youngkenny1996}, nonlinear modulational processes 
\cite{weatherall1998,cetal2001}, or driven thermal waves \cite{cetal2000}. 

   Pulsar variability and emission mechanisms can be addressed directly by 
analyzing the radiation statistics. Recently we presented a preliminary analysis 
of field statistics for the Vela pulsar \cite{cetal2001}, demonstrating 
that the observed variability is intrinsic and corresponds 
to lognormal statistics at some pulsar phases, interpreting these 
statistics and the observed variability in terms of a pure SGT system, so that 
the associated pulsar emission mechanism involves only 
linear processes. The overall goal of the present paper and its companion 
(Cairns et al., 2002, hereafter called Paper III) is to present a 
detailed analysis of the Vela pulsar's 
field statistics. Specific goals of this paper are: (i) to survey Vela's 
field statistics as 
a function of phase, showing evolution from approximately Gaussian to power-law to 
lognormal to power-law to Gaussian distributions as the pulse phase 
varies from off-pulse to on-pulse to off-pulse again, (ii) present 
single-component Gaussian 
and lognormal fits to the field statistics for appropriate pulsar phases, (iii)
demonstrate in detail that the variability corresponds to lognormal statistics 
near the peak of the pulse profile but Gaussian noise off-pulse, (iv) 
interpret the lognormal statistics in terms of SGT and linear emission mechanisms, 
and (v) discuss the results, including the constraints on nonlinear 
emission mechanisms. Paper III presents and 
interprets 
two-component fits to the observed field statistics over the remaining phase 
ranges; together these papers show that the Vela data are 
consistent with SGT applying, and with the 
emission mechanisms being purely linear, whenever the pulsar is detectable. 

     This paper proceeds by  
presenting required background information on theories for wave 
statistics (Section 2) and the Vela dataset (Section 3). Subsequently, the 
observed field statistics are surveyed, their evolution demonstrated, and 
initial analyses and associated interpretations described (Section 4). Detailed 
single-component fits to the observed field distributions for specific 
ranges of phase are then presented, first Gaussian intensity distributions 
off-pulse (Section 5) and then lognormal distributions near the center of the 
average pulse profile (Section 6). Brief descriptions of attempts to fit 
other distributions, specifically $\chi^{2}$ distributions and Gaussians in 
the field, to the data for these phase ranges are summarized in Section 7. 
Theoretical interpretations of these results are described and 
placed in context, first specifically for the Vela pulsar (Section 8) and then 
for other pulsars and astrophysical and solar system sources (Section 9). The 
conclusions are given in Section 10. 
 
\section{Theories for Wave Statistics}
 
   Wave growth in inhomogeneous plasmas naturally leads to bursty waves with 
widely varying fields. A qualitative rationale for this 
is as follows: (i) the plasma's inhomogeneity leads to localized regions being 
favored for wave growth (e.g., the plasma instability depends upon plasma 
density, temperature etc.), (ii) the distribution function $f({\bf v})$ for 
the driving particles relaxes more in the growth sites 
(leading to larger wave fields there), (iii) this injects spatially and 
temporally varying fluctuations into $f({\bf v})$, 
(iv) these fluctuations cause burstyness and variable waves 
along the particle path, and (v) the system evolves to a statistically steady state 
where possible. These wave-particle interactions are expected to drive the system towards 
marginal stability, where wave emission and damping (and related energy inflows and 
outflows) are balanced, time- and volume-averaged. One likely condition for 
reaching a steady state is that the 
unstable features in $f({\bf v})$ be able to reform, at least partially; e.g., due to 
fast particles outrunning slow particles in a beam system \cite{retal1993}. 
Different theories for wave growth are then characterized by different degrees of 
interaction between the waves, driving particles, and background plasma, 
leading to different wave statistics. Since these theories are not well known in 
the astrophysical literature, their salient features are summarized here in some 
detail. 

   SOC \cite{baketal1987,bak1996} is relevant to  
systems with fully self-consistent interactions between the waves, driving 
particles and background plasmas and with no preferred distance or time scales. 
It predicts power-law statistics for the distribution of energy releases 
(e.g., wave energies), so that the distribution $P(I)$ 
of wave intensities $I$ obeys (where $I \propto E^{2}$ for the wave electric 
field $E$)  
\begin{equation}
P(I) \propto I^{-\alpha} \ . 
\end{equation}
Typically the indices $\alpha$ approximately equal $1$, with a range $\approx 0.5 - 2$. 
Proposed examples include tokamak turbulence \cite{carrerasetal1996} and solar flares 
\cite{solar}. Also, 
Jovian ``S'' bursts have a power-law distribution for the radiation flux 
\cite{queinneczarka2001} and so may be an SOC system.

\begin{figure}
\psfig{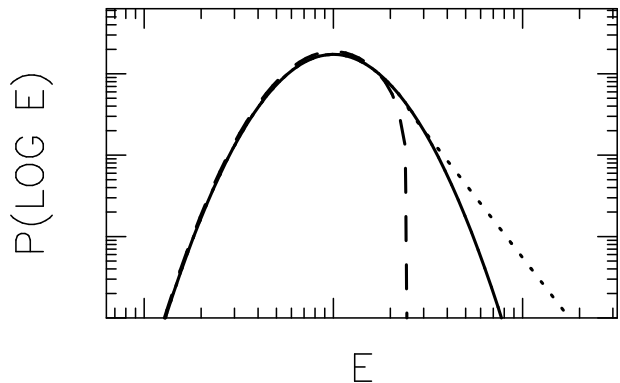}
\caption{Predictions for the $P(\log E)$ distribution for 
pure SGT (solid), SGT with a nonlinear decay process (dashed line), and 
SGT with wave collapse (dotted line). }
\label{fig2}
\end{figure}

   SGT (Robinson 1992, 1995l Robinson et al., 1993, 
Cairns et al. 2000, Robinson \& Cairns 2001, Cairns \& Menietti 2001) 
treats systems in which self-consistent wave-particle interactions occur in an 
independent, spatially inhomogeneous medium. The medium and wave-particle 
interactions then determine the relevant 
distance and time scales. In SGT the system evolves to a state in which (i) 
$f({\bf v})$ is close to time- and volume-averaged marginal stability but with 
fluctuations about this state which cause (ii) the wave gain $G$ to be stochastic 
variable. Relevant definitions are 
\begin{equation}
E^{2}(t) = E_{0}^{2} e^{G(t)}\ , 
\end{equation}
with $G = 2 \ln( E(t) / E_{0})$ for a reference field $E_{0}$, and  
\begin{equation}
G(t) = \int_{-\infty}^{t} dt\ \gamma(t) \label{G(t)}
\end{equation} 
where $\gamma(t)$ is the (amplitude) growth rate of the waves. SGT is then a 
natural theory for bursty waves with widely varying fields (due to $\log E \propto G$ 
being a Gaussian random variable) that persist with the driving particles for 
unexpectedly large distances and/or times (due to the closeness to marginal stability). 
Moreover, SGT should be widely applicable since there is a natural qualitative 
route to an SGT state: writing the time integral in (\ref{G(t)}) as 
\begin{equation}
\int_{-\infty}^{t} dt\ \gamma(t) = {\LARGE \Sigma}_{i} \Delta G_{i} \ , 
\end{equation}
then provided that sufficient fluctuations $\Delta G_{i}$ [associated with fluctuations 
in $f({\bf v})$] pass though a growth site during the characteristic time waves grow 
there, the Central Limit Theorem (CLT) predicts that $G$ will have Gaussian statistics 
irrespective of the detailed statistics of $\Delta G_{i}$. This 
means that $P(\log E) \propto P(G)$ obeys 
\begin{equation}
P(\log E) = (\sqrt{2 \pi} \sigma)^{-1}\ e^{- (\log E - \mu)^{2} / 2 \sigma^{2} )}\ , 
\label{p_sgt} 
\end{equation}
where $\mu = \langle \log E \rangle$ and $\sigma$ are the average and standard 
deviation of $\log E$, respectively, and $\log \equiv \log_{10}$. That is, pure 
SGT predicts lognormal statistics for the field (Figure \ref{fig2}). In recent years 
pure SGT has been 
shown to be widely applicable 
(Robinson et al. 1993, Cairns and Robinson 1997, 1999, Cairns et al. 2000, 
Cairns \& Grubits 2001, Cairns \& Menietti 2001, Paper I). 

  SGT can coexist at moderate $E$ with nonlinear processes active at high $E$, 
causing characteristic 
modifications to the predicted lognormal statistics at fields close to and 
above the threshold field $E_{c}$ for the active nonlinear process. The presence or 
absence of active nonlinear processes can thus be constrained using 
field statistics, as used successfully in several applications. Two types 
of modification are known, as illustrated in Figure \ref{fig2}. First, when a 
three-wave decay process takes energy 
out of the measured waves, the $P(\log E)$ distribution is cutoff at 
$E \gapprox E_{c}$ with known analytic form \cite{retal1993,cm2001,cg2001}. 
Second, the nonlinear process of wave collapse 
(Robinson 1997, Weatherall 1997, 1998),  
due to a self-focusing or modulational instability of a wavepacket, leads to a 
power-law tail at 
$E \gapprox E_{c}$. These indices are typically large: collapse of Langmuir waves 
in non-relativistic electron-proton plasmas  
leads to 
\begin{equation}
P(E) \propto  E^{-1} P(\log E) \propto E^{-\alpha} 
\end{equation} 
with $\alpha$ varying between $4$ and $7$, depending on the dimensionality (2-D to 3-D) 
and shape (isotropic versus oblate versus prolate) of the 
wave packets and whether the collapse is subsonic or supersonic \cite{rn1990,r1997}. 
The characteristics of collapse, or even whether it occurs, are not so 
clear for the strongly magnetized and relativistic electron-positron plasmas 
relevant to pulsar magnetospheres: while a 
purely electrostatic analysis \cite{pelletieretal1988} suggests that solitons 
are stable to modulational instability and do not collapse, electromagnetic 
simulations suggest instead that the solitons are subject to modulational instability 
and do collapse (Weatherall 1997, 1998). Although the field statistics of 
these collapse events are not known, the qualitative behaviour is very similar 
to that for collapse in electron-proton simulations (Weatherall 1997, 1998).  
Accordingly, it is presumed hereafter that, if 
modulational instability occurs, then it leads to wave collapse and power-law 
field statistics described by (6) with indices similar to those given above 
for electron-proton simulations. 
 
    
  SGT also describes the statistics of waves driven by a linear instability from 
the thermal level $E_{th}$ \cite{r1995,cetal2000}, being essentially a power-law with 
low index at fields 
$\gapprox E_{th}$, while thermal waves have a field distribution 
given by the product of a power-law with a Gaussian in $E^{2}$. 
In contrast, elementary burst theory [EBT] predicts exponential statistics 
(Melrose \& Dulk 1982, Robinson, Smith \& Winglee 1996) while
the formerly standard theory for wave growth \cite{kralltrivelpiece,melrose1986}, 
that of homogeneous exponential growth with constant 
growth rate until saturated at high fields by a nonlinear process or quasilinear 
relaxation (uniform secular growth), predicts $P(\log E)$ should be uniform (flat) 
below the nonlinear 
level \cite{retal1993,cr1999}. While EBT appears consistent with solar microwave 
spike bursts \cite{rsw1996,islikerbenz2001}, the statistics of all systems published 
to date are inconsistent with uniform secular growth. 

   The final model considered here for field statistics and variability is that 
of Gaussian statistics in $I$, i.e.,  
\begin{equation}
P(I) = (\sqrt{2\pi} \sigma_{I})^{-1}\ e^{- (I - I_{0})^{2} / 2 \sigma_{I}^{2} } \ , 
\label{p_i_eqn}
\end{equation}
where $I_{0} = \langle I \rangle$ and $\sigma_{I}$ are the average and standard 
deviation of $I$. This is the standard expectation for background noise, for instance 
due to superposition of multiple random signals, so that the sky background, measurement 
noise, and sources with multiple unresolved sources should have $P(I)$ 
distributions with Gaussian statistics. Scattering of waves between the source and 
observer, due to refraction by density irregularities, can also lead to Gaussian 
$P(I)$, as sketched next.  The 
Central Limit Theorem for a monochromatic (real, scalar transverse) field $E$ predicts 
Gaussian field statistics if sufficiently many ray paths contribute, whence the 
distribution $P(I)$ should then have an exponential distribution 
\cite{ratcliffe1956,rickett1977}, corresponding to a $\chi^{2}$ distribution with 
$2$ degrees of freedom. This distribution is well approximated by 
(\ref{p_i_eqn}) if the scattering is weak and/or $\sigma_{I} \lapprox I_{0}$ 
\cite{ratcliffe1956}. Intuitively, inclusion of finite bandwidth effects, due to multiple 
independent monochromatic signals contributing in a receiver bandwidth, should also 
lead towards Gaussian intensity statistics. 

   Relationships between the $P(I)$, $P(E)$ and $P(\log E)$ distributions follow 
from their normalization conditions, e.g., $\int dX\ P(X) = 1$, and differentials. 
Accordingly 
\begin{equation}
P(\log E) = \ln 10\ E\ P(E) = 2\ln 10\ E^{2}\ P(I)\ ,  
\end{equation}
allowing (1), (5), and (6) to be related quickly. 
In summary, the field statistics predicted by the above theories are known and 
can be compared robustly with observational data to  
(i) determine the relevance of particular theories for wave growth, thereby 
constraining the physics of the wave growth and source region and the causes 
for the source's variability, (ii) identify the presence or absence of 
nonlinear processes actively participating in the emission processes, and (iii) 
constrain the importance of scattering.

\section{Dataset}

   The dataset consists of $20,085$ contiguous intensity pulses of the Vela 
pulsar, spanning approximately $30$ minutes, measured at $1413$ MHz 
by the Parkes radio telescope and described in detail elsewhere 
\cite{johnstonetal2001}. Only a brief summary is given here. 
The measured quantities are the voltages (proportional to the received 
electric field strength) on two orthogonal feeds, which 
are frequency-downconverted, the DC offset removed with a time constant 
$\tau_{DC}$ long compared with the pulsar period ($\approx 88$ ms), and fed into 
a modified version of the Caltech and Princeton backend systems 
\cite{jenetetal1997,stairsetal2000}. The backend system then 
samples the datastream in quadrature with 2-bit accuracy at 20 MHz and writes it to 
digital linear tape. Offline, dispersion effects are removed using coherent 
dedispersion techniques, the four Stokes parameters are recovered, instrumental 
polarization effects removed, and a flux calibrator (Hydra A) used to 
write the output data in terms of flux densities $F$ averaged over the backend's 
20 MHz bandwidth. The resulting dataset has $2048$ phase bins per 
pulse period, each of $\sim 44~\mu$s length, which is comparable to the 
scatter-broadening time at this frequency. About $1800$ 2-bit samples are thus 
used to calculate the flux in each phase bin every pulse. Arguments against 
this digitizing procedure significantly modifying the true field statistics 
are presented in section 9. 

    The resulting brightness temperature $T_{br}(\phi)$ at a given pulse 
phase $\phi$, proportional to $F$ and ideally to the antenna's 
electric field squared, 
may be written as 
\begin{equation}
T_{br}(\phi) = T_{rec} + T_{snr} + T_{back} + T_{psr}(\phi) \ , 
\label{T_eqn}
\end{equation}
where $T_{rec}$ is the receiver (noise) temperature, $T_{snr}$ is the temperature 
of Vela's supernova remnant, $T_{back}$ is the background temperature due to 
galactic synchrotron radiation, the Sun, and other sources, and $T_{psr}$ 
is the pulsar's brightness temperature. Since the antenna points at the same location 
of the sky for the observing interval, removing the DC offset should eliminate 
$T_{snr}$ and $T_{back}$ from the final dataset (provided they do not vary on 
times $\ll \tau_{DC}$). Similarly, slow drifts in $T_{rec}$ will be removed, 
although rapid fluctuations corresponding to the receiver's thermal 
noise persist and naturally dominate the observed field statistics at off-pulse phases 
(see Section 5 below). 

    While removal of the slowly-varying DC offset is an advantage in terms 
of observing weak sources above receiver and sky backgrounds, it is also a 
disadvantage for analyses of field statistics. There are at least three reasons. 
First, since $E^{2} \propto I \propto T_{psr}$, neglecting the time-steady, 
phase-averaged 
part of $T_{psr}$ ($\langle T_{psr} \rangle$) affects the field scale nonlinearly 
for fields $E$ such that $T_{br}(\phi)$ is within a factor of a few times 
$\langle T_{psr} \rangle$, thereby potentially modifying the field statistics 
there. Second, the logarithim function, which features prominently 
in SGT and the observed field statistics shown below, is undefined for 
negative fluxes (which occur in the dataset due to removal of the DC offset and 
due to thermal fluctuations in the receiver noise).  Third, measurement of non-zero 
values of $\langle T_{psr} \rangle$ would 
permit more detailed investigation of the source physics, particularly if 
multiple wave populations are superposed, and perhaps allow 
competing physical models to be tested and distinguished. Possible origins for 
physically-significant, non-zero $\langle T_{psr} \rangle$ 
include synchrotron emission from the pulsar magnetosphere and 
coherently-produced radiation that undergoes scattering and diffusion as it 
propagates from its source, perhaps even changing the phase at which it is 
observed.  Since the values $\langle T_{psr} \rangle$ are not available for 
these data, these issues are addressed by adding a constant 
offset flux 
\begin{equation}
I_{off}' = 1250\ {\rm mJy} 
\label{offset}
\end{equation}
to each sample in the dataset. Clearly, 
when fitting Gaussian intensity distributions (\ref{p_i_eqn}) to the data, 
only relative variations in the centroid from $1250$ mJy are  
significant. 

   Comparisons between the data and Section 2's theories are most easily 
accomplished by recasting the data in terms of fields and intensities. This is 
not inappropriate anyway, since the radiation fields incident on the antenna 
were first detected as voltages and the final calibrated fluxes are averaged 
over the detector's $20$ MHz bandwidth. The calibrated intensity $I$ then equals the 
flux multiplied by the bandwidth and the field $E$ is proportional to $I^{1/2}$. 
Accordingly, rather than converting into 
SI units, the analyses below are cast in terms of variables $E'$ and $I'$, 
defined by 
\begin{eqnarray}
E'  = & ( F + 1250 {\rm mJy} )^{1/2} & \propto E \ , \\
I'  = & F + 1250 {\rm mJy} & \propto I \ , 
\end{eqnarray}
which are related directly to the calibrated fluxes $F$ and whose 
units are in (mJy)$^{1/2}$ and mJy, respectively. Accordingly, $E'$ and 
$I'$ differ by multiplicative constants from $E$ and $I$ 
(and ideally to the incident field and intensity also), respectively, 
which correspond to constant scale factors or offsets along the abscissa axis 
in the linear and logarithmic plots of field statistics 
below.

   Figure 1 shows the average pulse profile for relevant phase bins in 
mJy \cite{cetal2001}, together with three superposed 
pulses that illustrate the 
variability. Note that the noise level is very low compared with many 
earlier analyses, allowing investigation of the intrinsic 
field statistics. Additionally, the pulses are sampled rapidly in 
time, reaching the limits imposed by scatter-broadening. This 
permits detailed investigation of the field statistics as a function 
of pulsar phase. The significant variations in field statistics with 
phase shown below suggest that rapid sampling is important if 
accurate analyses of field statistics is desired. 

\section{Identification of Regions}

\begin{figure}
\psfig{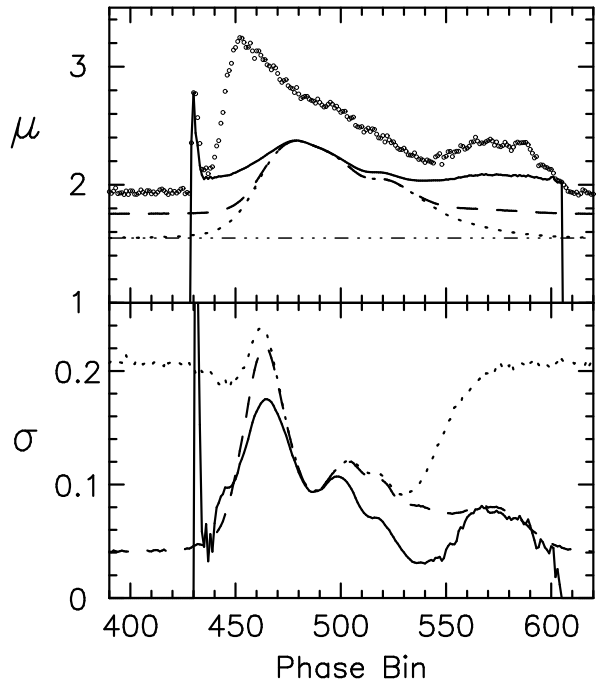}
\caption{(Top) The maximum value of $\log E'(\phi)$ (circles) and three 
estimates for the averages $\mu(\phi) = \langle \log E' \rangle (\phi)$, calculated using 
three $I'$  
thresholds (solid -- $10^{4}$ mJy, dashed -- $2.5\times 10^{3}$ mJy, 
and dotted -- 0 mJy), as functions of phase $\phi$ for the Vela dataset. 
(Bottom) Three estimates for $\sigma(\phi)$, using the line styles and 
intensity thresholds for (Top), as functions of phase. }
\label{mu_mumax_sigma}
\end{figure}

  The off-pulse phase bins considered (phases $390 - 420$ and $600 - 630$, cf. 
Figure \ref{fig1}) have average intensities $I' \approx I_{off}' 
= 1250$ mJy and maximum values $\approx 8000$ mJy.  Only the on-pulse bins 
have $I' \gapprox 10^{4}$ Jy. 
Figure \ref{mu_mumax_sigma} surveys the whole dataset as a function of 
$\phi$: the top panel shows the maximum value of $\log E'$ for each 
$\phi$, $\mu_{max}(\phi)$ (circles), and three averages for 
$\mu(\phi) = \langle \log E' \rangle (\phi)$ (solid, dashed, and dotted lines), 
corresponding to averaging $\log E'$ samples above specified intensity 
thresholds ($I' \ge 10^{4}$, $2500$, and $0$ mJy, respectively). The Figure's 
lower panel shows three estimates for $\sigma(\phi)$, which is the standard 
deviation of $\log E'$, for these same phases and thresholds. 

   The sharp, very localized peak in $\mu(\phi)$ and $\sigma(\phi)$ for a 
threshold of $10^{4}$ 
Jy at phases $429 - 433$, and its coincidence with the localized peak in $\mu_{max}$, 
demonstrates the appearance of a strongly phase-localized, intense component 
of the pulsar's output. This corresponds to the giant micropulses identified 
previously \cite{johnstonetal2001}. Otherwise Figure \ref{mu_mumax_sigma} 
shows smooth 
variations in $\mu$, $\mu_{max}$ and $\sigma$ with phase, including the ``bump'' 
identified for bins $550 - 600$ by Johnston et al. (2001). The effects of the 
different thresholds are not important for phases $\approx 460 - 540$, 
where the curves for the three thresholds agree well, corresponding to phases 
where $\mu(\phi) \gapprox 0.5 \log \sqrt{I_{off}}$ and the pulsar 
fields are typically well above the average value. However, the peaking of 
$\mu$ and $\mu_{max}$ at different phases ($\approx 480$ versus $\approx 450$) 
points towards different field statistics in these regions, as demonstrated 
in detail next, and to evolution in the number and characteristics of active 
components for the source.  

\begin{figure}
\psfig{file=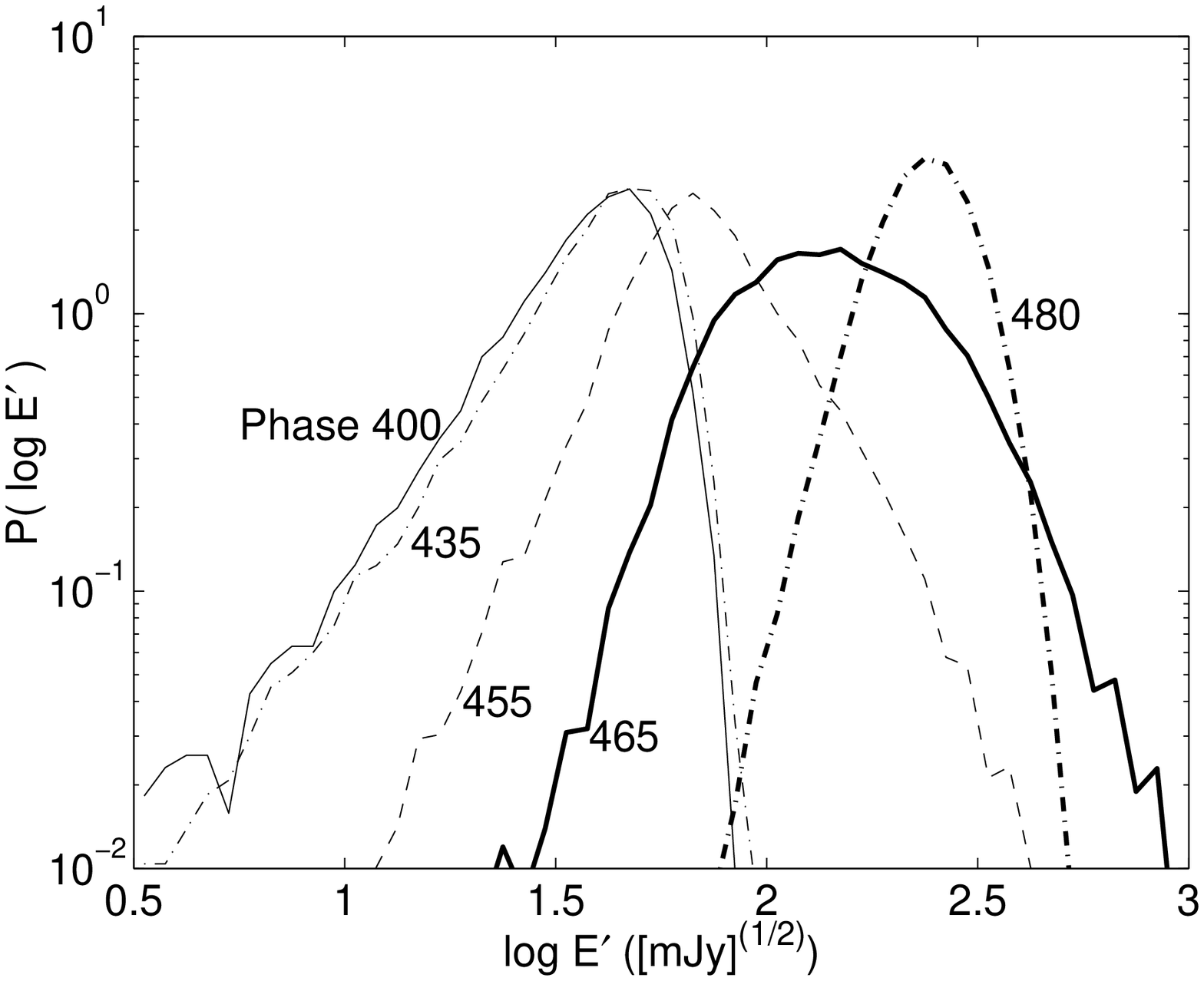,height=7.0cm}
\psfig{file=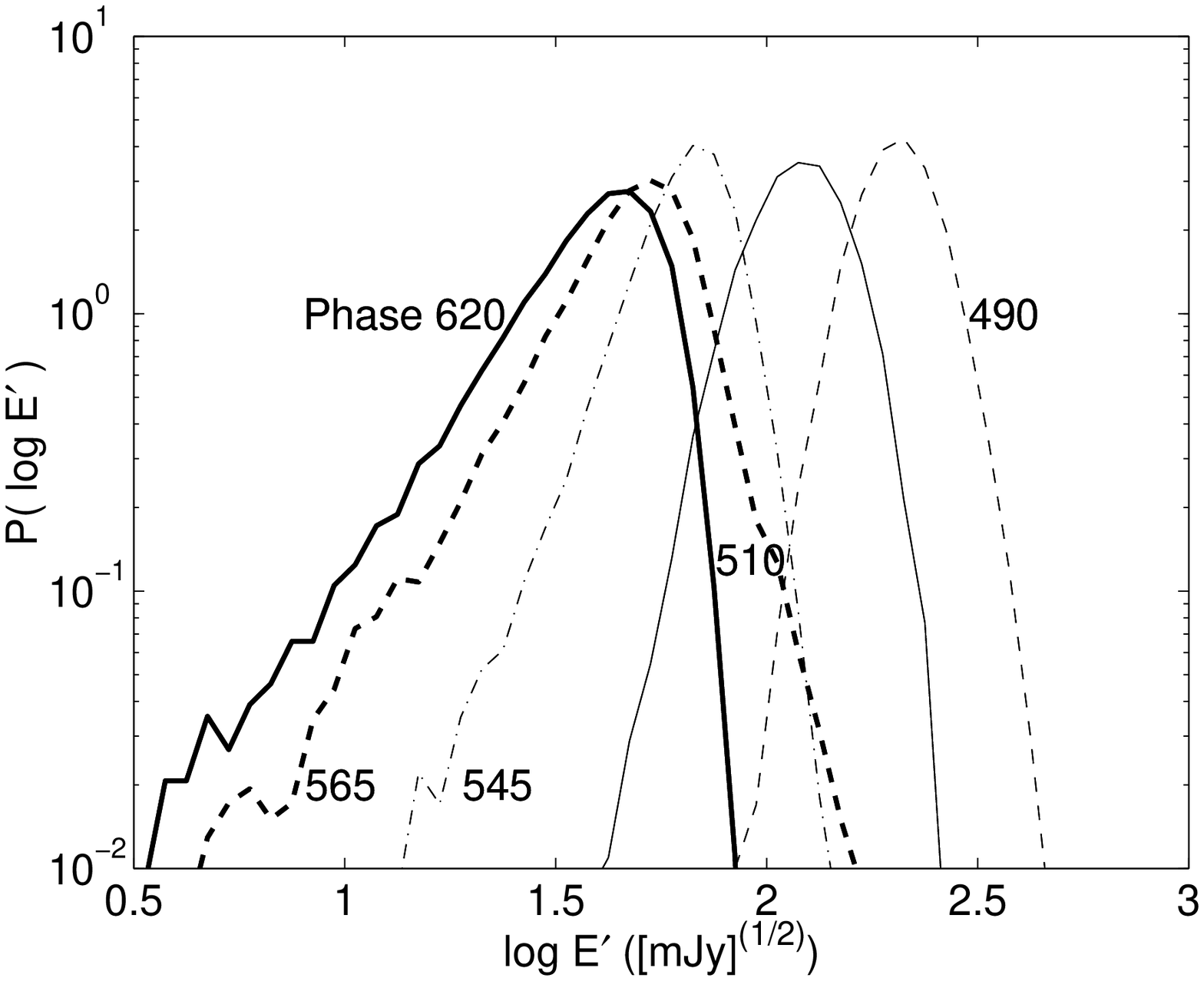,height=7.0cm}
\caption{Distributions $P(\log E')$ for chosen phases $\phi$, ranging 
from off-pulse ($\lapprox 430$) to pulse-center (e.g., 470 and 490) to off-pulse 
again ($\gapprox 600$).}
\label{p_loge_survey}
\end{figure}

   Figure \ref{p_loge_survey} shows the evolution of the field 
distribution $P(\log E')$ as a function 
of pulsar phase. At each phase this distribution is calculated by binning the 
intensity time series in $\log E'$ and normalizing so that 
$\int d(\log E') P(\log E') = 1$. 
The $P(\log E')$ distributions at phases $\lapprox 428$ and 
$\gapprox 600$ correspond to a Gaussian intensity distribution, being interpretable 
as Gaussian noise, as shown in detail in Section 5 below. However, for 
$\phi \gapprox 429$ first a localized 
high-$I$ tail appears and disappears [the giant micropulses \cite{johnstonetal2001}], 
after which an 
approximately power-law component appears at about phase $435$ and extends to 
increasingly high $E'$ until about phase $455$. After this the entire distribution 
moves towards higher $E'$, broadens, and the high-$E'$ power-law slope becomes 
approximately parabolic in $\log P(\log E') - \log E'$ space 
at phase 465, corresponding to a lognormal distribution. The peak of this
lognormal $P(\log E')$ distribution continues to move to higher $E'$ until 
about phase 480 but becomes narrower, corresponding to higher $\mu$ and smaller 
$\sigma$ in (\ref{p_sgt}). Subsequently, the distributions continue to be 
approximately parabolic 
but move to lower $\mu$ and $\sigma$. Eventually the effects of Gaussian intensity 
noise become evident, giving rise to a non-parabolic tail 
at low $E'$ for phases $\gapprox 540$. In the phase range $\approx 545 - 600$ a 
power-law tail at high $E'$ develops and then disappears, corresponding 
approximately to the ``bump'' \cite{johnstonetal2001} in $\mu$ in 
Figure \ref{mu_mumax_sigma}, after which the 
field distribution recovers the form in the off-pulse bins below phase $428$. 
 
     Figure \ref{p_loge_survey} thus directly demonstrates that the Vela 
pulsar has different field statistics in neighboring ranges of phases; put another 
way, the field statistics vary with $\phi$ across the source. This immediately 
shows that field statistics can be used to probe the source physics. 

\begin{figure}
\psfig{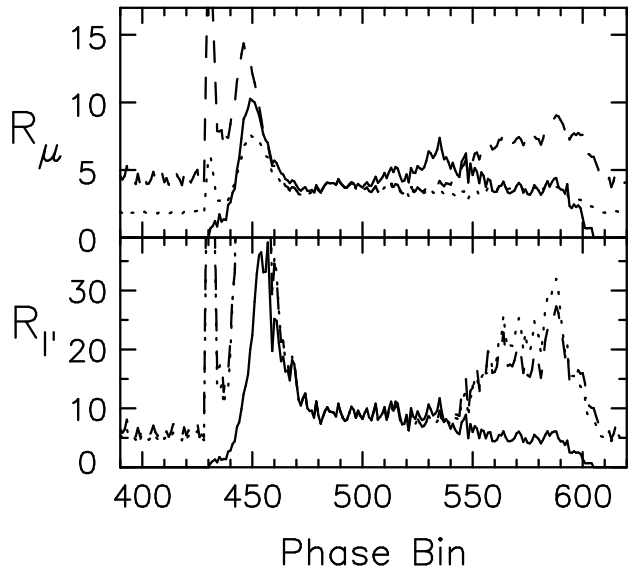}
\caption{The quantities $R_{I'}$ and $R_{\mu}$ given by (13) and (14), 
respectively, are shown as functions of phase using the same intensity 
thresholds and line styles as Figure \ref{mu_mumax_sigma}.} 
\label{R_fig}
\end{figure}

The qualitative identifications given above are elaborated next by 
considering the quantities 
 \begin{eqnarray}
 R_{I}(\phi) & = & [ I_{max}'(\phi) - \langle I' \rangle (\phi) ]\ /\ \sigma_{I}(\phi) \\
 R_{\mu}(\phi) & = & [ \mu_{max}(\phi) - \mu(\phi) ]\ /\ \sigma(\phi) \ , 
 \end{eqnarray} 
the first of which was considered previously, for example, by Johnston 
et al. (2001).   
 A Gaussian intensity distribution at phase $\phi$ is expected to have 
 $R_{I'}(\phi) \lapprox 5$, since almost all its samples should be within 
 $\lapprox 5\sigma_{I'}(\phi)$ of the mean $\langle I' \rangle(\phi)$. A similar 
 statement follows for a lognormal field distribution, which should have 
 $R_{\mu}(\phi) \lapprox 5$. Figure \ref{R_fig} shows these quantities at different 
 phases for the three intensity thresholds used in Figure \ref{mu_mumax_sigma}. 
 The $R_{I'}(\phi)$ 
 values in Figure \ref{R_fig}'s top panel show that Vela's field statistics 
 are close to Gaussian in $I'$ only for phases $\lapprox 428$ and 
 $\gapprox 610$. The $R_{\mu}$ values in Figure \ref{R_fig}'s bottom panel 
 show that 
 the field statistics are close to lognormal at phases $\gapprox 460$ and 
 $\lapprox 540$, with particularly close agreement evident in the range $460 - 510$ 
 where the traces for all three threshold are very close together. 
 
    Figures \ref{mu_mumax_sigma} to \ref{R_fig} are thus all qualitatively 
consistent with the following 
identifications: phases $\lapprox 428$ and $\gapprox 600$ 
have Gaussian intensity statistics \cite{cetal2001}, phases $429 - 433$ 
have non-Gaussian and 
non-lognormal statistics that correspond to giant 
micropulses \cite{johnstonetal2001,krameretal2002}, phases with $435 - 455$ 
and $545 - 600$ have an approximately power-law
character at large $I'$ and $\log E'$; phases $460 - 540$ have lognormal 
field statistics (Papers I and III). These 
identifications are quantified next, using single-component fits,  
for the Gaussian-$I'$ and 
lognormal regimes, refining and extending the analysis of Paper I. 
The approximately power-law domains are termed the ``transition region'' in Paper 
III, since these distributions are best interpreted in terms 
of vector convolution of a Gaussian and a lognormal component and not in terms 
of intrinsic power-law statistics. 
Two-component fits for the transition region and lognormal domains are detailed in  
Paper III. 
 
\section{Gaussian noise region}
 
\begin{figure}
\psfig{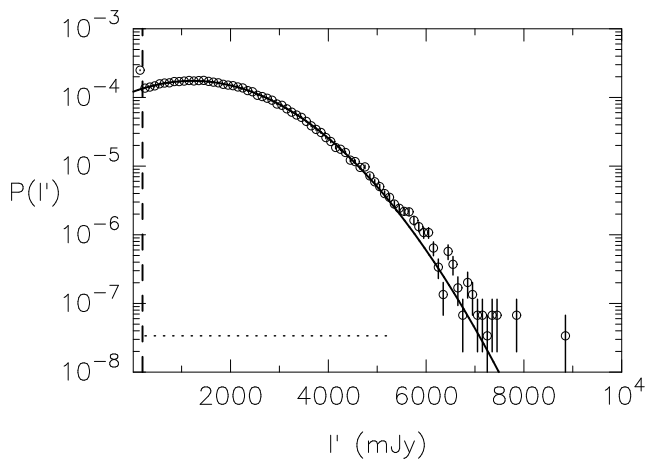}
\caption{Comparison of the observed intensity distribution $P(I')$ for 
phases $391-399$ (circle symbols with error bars given by counting 
statistics) with the best fit to the prediction (\ref{p_i_eqn}) for 
Gaussian statistics (solid line). Fits are performed where the dotted 
line shows that bins have $\ge 100$ counts and to the right of the 
vertical dashed line at $I' = 200$ mJy.}
\label{p_i}
\end{figure}

    The field statistics at phases 
$\lapprox 428$ and $\gapprox 600$ are now shown in detail to be Gaussian in 
the intensity $I'$, as expected for instrumental 
noise, background ``noise'' formed by incoherent superposition of a large 
number of small signals, and/or scattering. Figure \ref{p_i} shows the 
$P(I')$ distribution 
observed for phases $391 - 400$, calculated by binning the data into 
linear intensity bins and normalizing, together with the corresponding fit  
to (\ref{p_i_eqn}), obtained using the Amoeba algorithm to minimize 
$\chi^{2}$ (Press et al. 1986). Restricting the fit to intensity bins with 
at least $100$ samples, the fit parameters are $\langle I' \rangle = 1220$ mJy 
(agreeing 
with $I_{off}'$ to within less than the $100$ mJy bin width), $\sigma_{I'} = 1420$ mJy, 
$\chi^{2} = 67$ for $N_{dof} = 47$ degrees of freedom, and a significance 
probability $P(\chi^{2}) = 0.03$. Clearly the fit (\ref{p_i_eqn}) agrees well with 
the observed data and has reasonable statistical significance. 

    At high $I' \gapprox 7000$ mJy the observed distribution hints at deviations 
from the Gaussian form (\ref{p_i_eqn}). The simplest explanation is that these 
high-$I'$ samples 
correspond to pulsar emissions that slightly exceed the background. Put another way, 
the phase range 391-400 is not entirely off-pulse. 

   Lognormal fits to Figure \ref{p_i}'s data are clearly inferior (not shown). This 
is expected based on the strong 
dependence of the $R_{\mu}(\phi)$ curves in Figure \ref{R_fig} on the intensity 
thresholds used, 
in contrast to the lack of variation of the $R_{I'}(\phi)$ curves on 
these thresholds. Thus these data are consistent with Gaussian intensity distributions 
and not lognormal field (or intensity) distributions. Analyses of other phases in the domain 
stated above yield similar fit parameters and statistics. This is expected on the 
basis of the very similar $P(\log E')$ distributions shown in Figure
\ref{p_loge_survey}.

\section{Pure Lognormal Region} 

   Single-component lognormal fits are presented in section 6.1 for the 
on-pulse phase domain $470 - 540$. Suggestions that a second component may 
also contribute are discussed in section 6.2. Two-component fits, 
which lead to very good agreement between observation and theory over 
the entire phase range, are presented in Paper III. 
 
\subsection{Single component fits}

\begin{figure}
\psfig{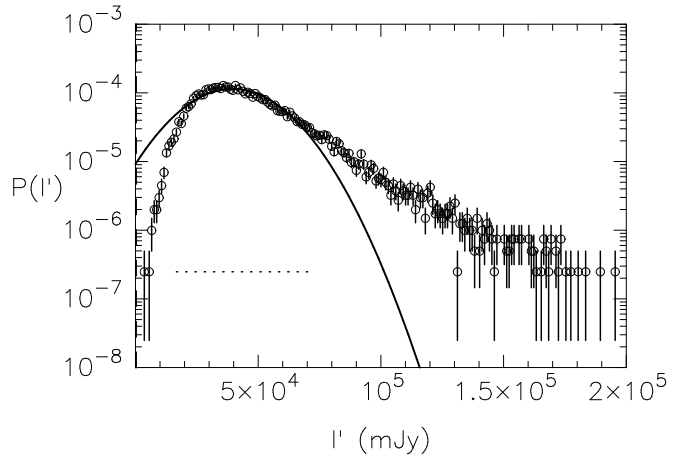}
\caption{Comparison of the intensity distribution $P(I')$ observed 
at phase 490 with 
the best fit (\ref{p_i_eqn}) for Gaussian statistics (solid line). The circle symbols, 
error bars, dotted line, and vertical dashed line are as in Figure \ref{p_i}. }
\label{p_i_490}
\end{figure}

   Figure \ref{p_i_490} shows the $P(I')$ distribution and its best Gaussian fit 
for phase 490, 
close to but after the peak in the pulsar's average profile (Figure 1). The fit 
clearly 
fails at both low and high $I'$, entirely missing the long tail at large $I'$,  
and has 
$\chi^{2} = 301$ for $N_{dof} = 53$ and $P(\chi^{2}) < 10^{-36}$ for the 
fitted range of $I'$ (dotted horizontal line). The 
variability at this phase is thus not described by Gaussian intensity statistics. 

\begin{figure}
\psfig{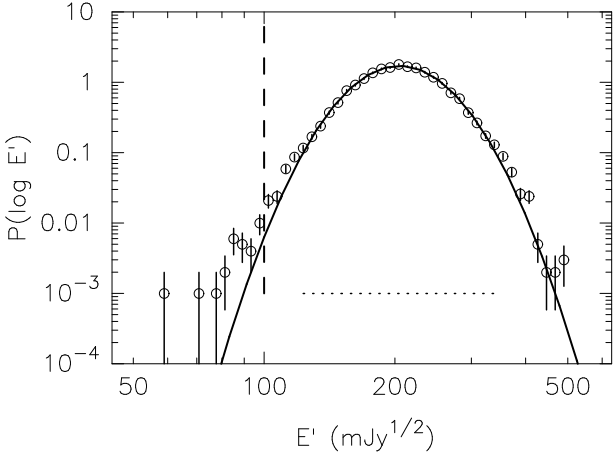}
\caption{Comparison of the distribution $P(\log E)$ observed 
at phase 490 with the best fit (\ref{p_sgt}) for lognormal 
statistics (solid line). The format is identical to Figure \ref{p_i} 
except that the vertical dashed line is now at $E' = 100$ mJy$^{1/2}$. }
\label{p_loge_490}
\end{figure}

   In contrast, Figure \ref{p_loge_490} shows that the $P(\log E')$ distribution 
for phase 490 is 
well fitted by the pure SGT prediction (5): for bins with $\ge 100$ counts (dotted line) 
and $E' \ge 10^{2}$ (mJy)$^1/2$ ($I' \ge 10^{4}$ mJy, which is $6\sigma_{I}$ above 
$\langle I' \rangle$ in Figure 6), the fit parameters are $\mu = 2.3$, $\sigma = 0.096$, 
$\chi^{2} = 27$  for $N_{dof} = 19$ and $P(\chi^{2}) = 0.12$. Similarly the 
Kolmogorov-Smirnov test (Press et al. 1986) yields a significance probability 
of $0.47$. This fit is strongly statistically significant:  
pulsar variability at this phase is lognormally distributed and quantitatively 
consistent with the theoretical form (\ref{p_sgt}) predicted for pure SGT. Note that 
the fit matches the data well even outside the fitted range of fields (vertical dashed 
line and horizontal dotted line), although the effects of the noise background become 
increasingly evident at fields $\le 80$ (mJy)$^{1/2}$. It is worth emphasizing that the 
 $P(I')$ distribution in Figure \ref{p_i_490}  
for $3\times 10^{4} \lapprox I' \lapprox 1.7 \times 10^{5}$ mJy does not have 
an intrinsically power-law form but is instead best represented in terms of a 
lognormal form (Figure \ref{p_loge_490}).  

\begin{figure}
\psfig{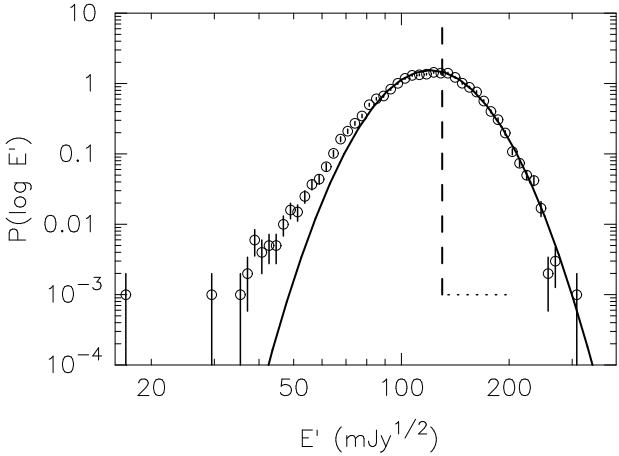}
\caption{Comparison of the distribution $P(\log E')$ observed 
at phase 510, in the same format as Figure 8, with the best fit
to the prediction (\ref{p_sgt}) for lognormal 
statistics (solid line). }
\label{p_loge_510}
\end{figure}

    Figure \ref{p_loge_510}  presents the observed 
$P(\log E')$ distribution and  
fit to the pure SGT prediction (\ref{p_sgt}) for phase 510 . 
Again the observed distribution is well fitted by 
pure SGT with good statistical significance. For fields 
$E' \ge 130$ (mJy)$^{1/2}$ the fit parameters are 
$\mu = 2.1 \pm 0.1$, $\sigma = 0.10 \pm 0.01$, 
$\chi^{2} = 13$ for $N_{dof} = 6$ with $P(\chi^{2}) = 0.04$. The statistical 
significance changes somewhat with the fitting threshold in $E'$: requiring 
$E' \ge 10^{2}$ (mJy)$^{1/2}$ yields $\mu = 2.1 \pm 0.1$, $\sigma = 0.11 \pm 0.1$, 
$\chi^{2} = 42$, $N_{dof} = 12$ and $P(\chi^{2}) = 3 \times 10^{-5}$, while 
requiring $E' \ge 150$ (mJy)$^{1/2}$ yields $\mu = 2.1 \pm 0.1$ and 
$\sigma = 0.10 \pm 0.01$ but $\chi^{2} = 7$ 
for $N_{dof} = 3$ and $P(\chi^{2}) = 0.07$. These varying statistical  
significances correspond to the varying contribution of the noise background 
to the observed distribution, due to the relatively weak pulsar fields at this 
phase and to $\mu$ approaching the background noise level. In each case the form 
of the distribution at high $\log E'$ is very well fitted by the SGT prediction. 

\begin{figure}
\psfig{file=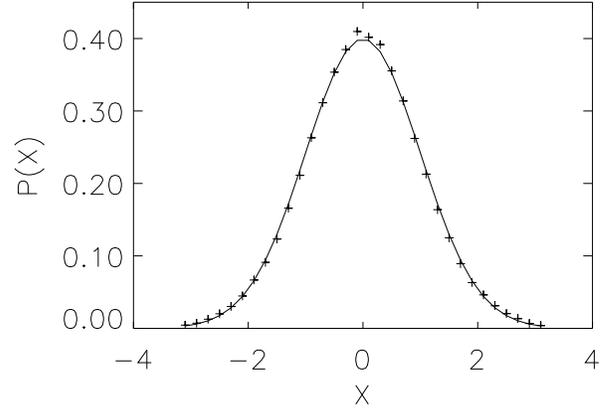,angle=0,height=6.5cm}
\caption{Comparison of the distribution $P(X)$ for all data in 
phases 485--495 (plus signs), calculated using 
the procedure described in the text, with the pure SGT prediction 
(solid line). }
\label{p_x}
\end{figure}

    Results similar to Figure 8 -- 9 are found for phases $470 - 540$, 
although the statistical significance varies. Rather than showing more results 
for individual phases, Figure \ref{p_x} shows the distribution $P(X)$ observed for 
phases $485 - 495$ \cite{cetal2001}. Here 
$X = [ \log E' - \mu(\phi) ] / \sigma(\phi)$ 
is the field variable resulting from detrending variations in $\mu$ and $\phi$ 
with phase $\phi$. Comparison with  (\ref{p_sgt}) shows that purely linear SGT predicts 
the $P(X)$ distribution to be Gaussian with zero mean and unit 
standard deviation \cite{cr1999}. Figure \ref{p_x}
demonstrates that the observations and SGT prediction agree well. 
The statistical significance is also good, with the Kolmogorov-Smirnov test 
yielding a significance probability of $0.1\%$. 

\begin{figure}
\psfig{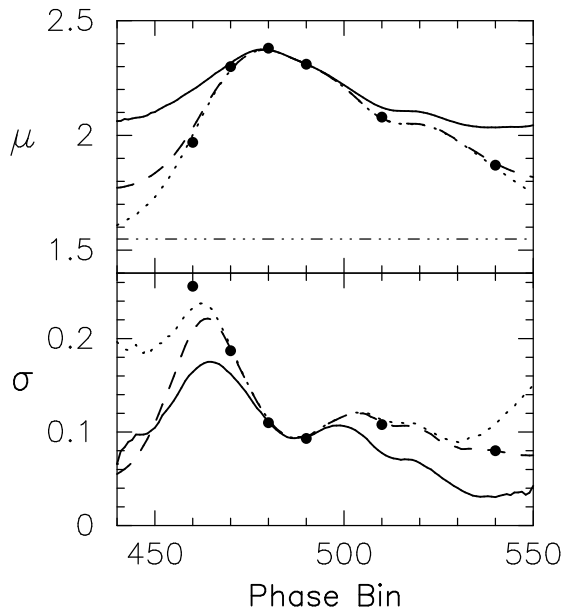}
\caption{Comparison of $\mu$ and $\sigma$ (circle symbols) obtained by fitting 
the $P(\log E')$ 
distributions to the SGT prediction (\ref{p_sgt}) with the values obtained 
directly from the data set for $I' \ge 2500$ mJy in Figure \ref{mu_mumax_sigma}. }
\label{mu_sigma_comp}
\end{figure}

   One test of the results of Figures 8 -- 10 is to compare the fits for 
$\mu$ and $\sigma$ with Figure \ref{mu_mumax_sigma}'s values calculated directly 
from the data. Figure \ref{mu_sigma_comp} shows that 
these values agree well, confirming that the fits are accurate and the 
lognormal component dominates the field statistics at these phases.   

\subsection{Initial evidence for a second component}

   Figures 8 -- 10 show that the pure lognormal form (\ref{p_sgt}) fits 
the data very well above the peak in the $P(\log E')$ distribution 
but less well at low $E'$. Specifically the observed distribution lies 
above the fit to (\ref{p_sgt}) at low $E'$, suggesting a contribution there from a 
second group (or component) of waves. This is not unexpected in view of the 
Gaussian statistics found at off-pulse phases (e.g., Figure \ref{p_i}), 
which if it corresponds to measurement noise, scattered radiation, or 
contributions from multiple unresolved, incoherently summed sources, might 
be expected to contribute at all phases. 

   In Paper III it is 
shown that the off-pulse Gaussian ``noise'' in Figure \ref{p_i} indeed 
persists into the on-pulse bins and perhaps evolves into a second lognormal 
component. Performing two-component fits to the observed $P(\log E')$ 
distributions then leads to very good agreement between observation and 
theory over almost the entire range of $E'$ for all on-pulse bins. 
In particular, the approximately power-law $P(\log E')$ distributions in 
Figure \ref{p_loge_survey} correspond to vector convolution of the Gaussian component 
with an emerging lognormal component, while the clearly lognormal 
distributions observed at phases 460 -- 540 are best modelled in 
terms of convolution of two lognormal components. 

\section{Attempts to Fit Other Distribution Functions}

   Attempts were made to assess the uniqueness of the Gaussian and lognormal fits 
presented heretofore, by also considering several other fitting functions: a Gaussian 
in the field (rather than the intensity), the SGT prediction for a thermal 
wave distribution \cite{r1995,cetal2000}, and a $\chi^{2}$ distribution in $I'$. Both the 
first and last of these are related to scattering theory \cite{ratcliffe1956,rickett1977}, 
the first being predicted for a scattered monochromatic real field and the latter 
(with $2$ degrees of freedom) for the intensity of a monochromatic scattered field. 
In more detail, a $\chi^{2}$ distribution with $n$ degrees of freedom corresponds 
to the distribution expected from summing a limited number $n$ of Gaussian-distributed 
variables and it is defined by 
\begin{equation}
P_{\chi}(n,I') = (I' / I_{1})^{n/2 - 1} e^{-I' / 2 I_{1}} / 2^{n/2} \Gamma(n/2) \ . 
\label{chi_squared_distrib}
\end{equation}
Here $I_{1}$ is related to the average $\langle I' \rangle$ by 
$I_{1} = \langle I' \rangle / (n + 2)$. In the limit $n \rightarrow \infty$ a 
$\chi^{2}$ distribution tends to a Gaussian. 

    Applying these three alternatives led to unsatisfactory 
results for all off-pulse phase bins (not shown), both in absolute terms and 
relative to the Gaussian intensity fits described in Section 5. This implies that 
the off-pulse data are best modelled in terms of Gaussian intensity statistics. 

\begin{figure}
\psfig{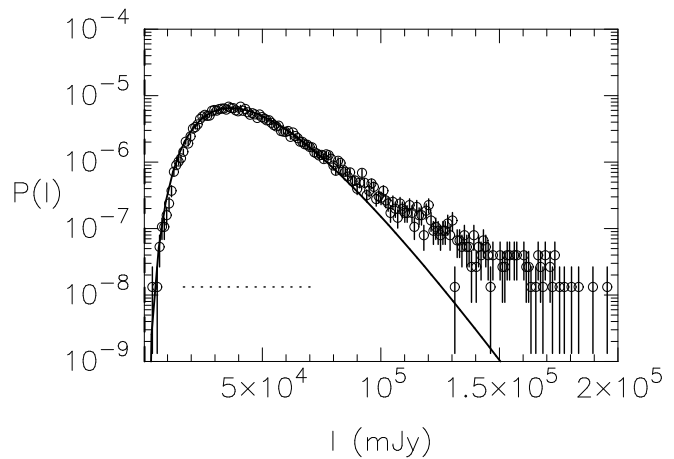}
\caption{Comparison of the distribution $P(I')$ observed 
at phase 490 with the best fit to prediction (\ref{chi_squared_distrib})
 for a chi-squared distribution, 
in the same format as Figure \ref{p_i}. }
\label{p_i_chisq}
\end{figure}

    The results of these fits for on-pulse bins in the range $470 - 540$ were 
also unsatisfactory, both in absolute terms and relative to the lognormal 
fits in Section 6. Specifically: (i) These data were 
not well fitted as being Gaussian distributed in $E'$ or as a thermal SGT 
distribution for any phase. (ii) Fits to the chi-squared 
distribution (\ref{chi_squared_distrib}) invariably failed at high $\log E'$ 
where the lognormal fits were clearly superior, due to an inability to fit 
the high-$I'$ tail. (iii) While the 
data could be well fitted by (\ref{chi_squared_distrib}) for some 
phase bins, but not all, the fit parameters $n$ and $I_{1}$ varied 
widely rapidly from phase bin to phase bin. (iv) Both the statistical significance 
of the fits and the size of the domains in $\log E'$ for which good agreement 
existed between the $P(\log E')$ 
data and fitting curves were generally considerably larger for the lognormal form 
(\ref{p_sgt}) than for the chi-squared distribution (\ref{chi_squared_distrib}). These 
statements are illustrated in Figure \ref{p_i_chisq} for phase 490: for this phase (but 
not all) there is very good agreement at low $I'$, as well as the typical inability 
to fit the high-$I'$ tail. Moreover, the 
fit to (\ref{chi_squared_distrib}) for phase 490 and bins with in excess of 100 
samples has $n = 12.3$, $I_{1} = 3560$, $N_{dof} = 54$, 
$\chi^{2} = 93$ and $P(\chi^{2}) = 8\times 10^{-4}$, cf. the results in 
section 6, while the best fit for phase 510  has $n = 8.3$,  
$I_{1} = 1860$ mJy, and $P(\chi^{2}) = 0.02$. 
Accordingly, it 
is concluded that the on-pulse data are not well described in terms 
of Gaussian field statistics, thermal SGT, or a $\chi^{2}$ distribution in $I'$, but 
are instead best described (for the four fitting functions attempted here) 
as lognormally distributed. 

\section{Theoretical Implications for the Vela Pulsar}

   In general, the observed field statistics are determined by intrinsic field 
statistics produced by relevant generation mechanisms, possible spatial 
variations across the source, and propagation effects. Propagation effects are 
discounted in the present context, since weak or moderate scattering of radiation 
by density fluctuations is typically predicted to cause Gaussian or exponential 
statistics of $I'$ \cite{ratcliffe1956,rickett1977}, not lognormal statistics. 
In this dataset, 
then, Vela's variability is not due to scattering by density irregularities. In 
general, though, scattering is expected to play some role in determining the 
field statistics of pulsars.
The simplest interpretation is therefore adopted now, that the field statistics 
observed at each phase are intrinsic and are not significantly affected by 
spatial variations in the source. The possibility of two generation 
mechanisms being active at a given phase is deferred to Paper III. 

   For the phase range 470 - 540 considered above, where the Vela pulsar's average 
pulse profile is well above the noise, the field distributions do not have 
power-law tails or nonlinear cutoffs but instead are consistent with 
lognormal statistics. Accordingly, these data are consistent with the 
observed variability being a direct manifestation of a simple 
SGT state, with no evidence for SOC or uniform 
secular growth.  Put another way, Vela's variability is due to lognormal 
field statistics and is consistent with the emission taking place in 
a source plasma that is in a simple SGT state. 

   The simplest interpretation is, then, that the absence of a power-law tail 
or cutoff in the $P(\log E')$ 
distributions for these phases is inconsistent with nonlinear processes 
(e.g., Pelletier et al. 1988, Asseo et al. 1990, Asseo 1996, Weatherall 1997, 1998) 
such as wave collapse, modulational instability, and three-wave decay processes 
playing a significant role. Instead, the lognormal statistics observed 
are consistent with linear emission mechanisms, 
implying that a plasma instability in a SGT state either directly generates 
the radiation or else generates non-escaping waves that are transformed 
into escaping radiation by linear processes (e.g., mode conversion) alone. 
Moreover, these results are consistent with recent conclusions 
\cite{melrosegedalin1999} 
that linear mechanisms are favoured on theoretical grounds and are 
made more topical by recent proposals of new linear emission mechanisms 
\cite{gedalinetal2002}. 

   At a deeper level, however, nonlinear mechanisms cannot be ruled out 
entirely. Instead, stringent conditions are imposed: a nonlinear 
mechanism is viable only if it produces lognormal statistics when averaged over 
a suitable ensemble of nonlinear wavepackets or structures, with no evidence 
for a power-law tail or cutoff. Since existing simulations of collapse yield 
power-law statistics when ensemble-averaged in a homogeneous system 
\cite{rn1990,r1997}, mechanisms involving collapse are not likely to be viable. 

   From the definitions of $\mu$ and $\sigma$ and the intensity decreasing 
with distance $R$ as $R^{-2}$, it is easy to show that $\sigma(R)$ is  
constant and that $\mu(R) = \mu(R_{0}) - \log(R/R_{0})$, 
where $\mu(R_{0})$ is the value at the source's edge ($R = R_{0}$). 
Taking the values $\mu = 2.0$ and $\sigma = 0.1$ to be representative 
of these phases, the distance $R = 350$ pc for Vela, and the 
value $R_{0} = 30$ m, yields $\mu(R_{0}) \approx 20$. 
The value $R_{0} = 30$ m results from assuming that the overall 
source is annular, with radius equal to the neutron star radius $\approx 
10$ km, and dividing by the $2048$ phase bins used for Vela. Accordingly, the ratio 
$\mu_{0}/\sigma \approx 200$ in the source. The values $\mu_{0}$ and 
$\sigma$ will constrain future theoretical models for why SGT applies, 
similar to existing models for solar system phenomena \cite{retal1993,cr1999,cm2001}. 

\section{Discussion}

   The foregoing analyses are the first detailed applications of SGT to propagating 
EM radiation and, simultaneously, to extra-solar system sources (see also 
Paper I). Their 
success implies that radiation statistics are an underappreciated and 
potentially very powerful tool in astrophysics (and space physics), and 
suggests that SGT may well be widely applicable to coherent astrophysical 
sources. Of course, SGT is not likely applicable to all sources or indeed to all 
components of pulsar emissions, as discussed further in Paper III. 

    The importance of field statistics as a probe of source physics is shown 
above in terms of emission mechanisms. This is also shown empirically in 
terms of source structure by the demonstrated evolution in field statistics 
as a function of 
Vela's phase, in particular by the field statistics ranging from Gaussian in 
$I$ off-pulse to approximately power-law, lognormal, and power-law on-pulse and to Gaussian 
in $I$ off-pulse again. This is the first detailed investigation of field 
statistics for a pulsar and the first demonstrated evolution in field statistics 
across a source. 
Quantitative discussion of how the fit 
parameters $\mu$, $\sigma$ etc. of the field statistics vary across Vela are 
deferred \cite{cetal2002a}, as is discussion of how representative 
the Vela results are to other pulsars.

   It is appropriate to discuss the possibility that the 
intrinsic field statistics are modified by the digitization, summation, and 
coherent dedispersion procedures associated with the telescope backend system and 
consequent processing of the dataset (see section 3). Four observational 
results that argue against this possibility are the following: (i) the 
off-pulse statistics 
are Gaussian in $I'$, as expected for receiver thermal noise; (ii) the on-pulse 
field statistics evolve smoothly with phase and the pulse profile (see section 4) 
and have well-defined functional forms; 
(iii) power-law field statistics are obtained from this dataset for Vela's giant 
micropulses \cite{krameretal2002} with indices of order those observed for 
giant pulses from other pulsars 
\cite{cognardetal1996,johnstonromani2001,romanijohnston2001,c2002}; (iv) 
the field distributions observed on-pulse can 
be fitted very well with lognormal functions \cite{cetal2001} and combinations 
of lognormal and Gaussian functions \cite{cetal2002a}, and can be interpreted theoretically 
in terms of existing theories for wave growth in plasmas; 
None of these results are expected {\it a priori} if measurement and analysis 
techniques have modified the intrinsic statistics significantly. Moreover, 
result (iii) is consistent either with the backend system and processing 
system not modifying the intrinsic field statistics or else with all analyses of 
giant pulses being 
similarly flawed.  Thus, while modification of intrinsic field statistics by the 
measuring process is possible and must be kept in mind, the above points are 
strong arguments against Vela's intrinsic field statistics being significantly 
modified for this dataset. 

   One final remark before concluding is that detailed analyses of field 
statistics would benefit from estimates for $\langle T_{psr} \rangle$ 
in (\ref{T_eqn}), corresponding to the value $I_{off}'$ in (\ref{offset}). 
This would allow an absolute scale for $E'$ and $I'$ to be estimated accurately 
and aid in interpreting detailed fits at low $E'$ of the field statistics 
\cite{cetal2002a}. 
 
\section{Conclusions}

  Analysis of rapidly-sampled, coherently dedispersed data show for the 
first time that Vela pulsar's variability 
corresponds to field statistics that are (i) well defined and (ii) evolve 
smoothly from 
Gaussian intensity statistics off-pulse to power-law  and lognormal statistics 
on-pulse. Since theories for wave growth in inhomogeneous plasmas predict the 
field statistics, these observations allow the source plasma and emission physics 
to be probed. Detailed 
single-component fits to the observed field statistics confirm that the off-pulse 
variability corresponds to Gaussian intensity statistics, consistent with superposition 
of multiple incoherent signals and/or scattering, while the lognormal statistics 
observed near the peak of Vela pulsar's average profile are 
consistent with the predictions of SGT 
for a purely linear system near marginal stability. The simplest interpretations 
are that the Vela pulsar's on-pulse variability is  
a direct manifestation of an SGT state and that only linear emission 
mechanisms (either direct or indirect) are viable. This argues against 
some nonlinear theories for pulsar radio emissions. At a deeper level nonlinear 
mechanisms are not ruled out but are strongly constrained: viable mechanisms 
must produce 
lognormal statistics when suitably ensemble-averaged. Accordingly mechanisms 
involving wave collapse do not appear viable due to them yielding power-law 
statistics. The data suggest that scattering 
minimally affects the field statistics at on-pulse phases for Vela. 
Two-component fits for Vela's power-law and 
lognormal domains, described elsewhere \cite{cetal2002a}, extend and strengthen 
these results, finding that the observed variability is consistent with lognormal 
statistics whenever the average pulse profile is above background. 

These analyses of pulsar emissions represent the first applications 
of SGT to both non-solar-system sources and to propagating, free-space radio 
emissions. This generalizes the theory's applicability significantly beyond 
the relatively localized, solar system, plasma waves considered previously. While 
SGT thus 
applies in all $8$ analyses performed by us to date, other field statistics 
are sometimes observed for other wave phenomena. Analysis of field statistics 
is thus a powerful 
tool for understanding source variability and constraining emission 
mechanisms and source characteristics. This new window into the source physics is 
likely to be widely useful for coherent astrophysical and solar system radio 
emissions, as already found for plasma waves in space. 

\section*{Acknowledgments}

The authors acknowledge financial support from the Australian Research Council 
and the University of Sydney, and helpful conversations with B.~J. Rickett, P.~A. 
Robinson, D.~B. Melrose, and Q. Luo.

\bibliographystyle{mn}
\appendix

\bsp

\label{lastpage}

\end{document}